# DRAMATURGIES OF DECEPTION: AI HUMANIZERS AND THE PERFORMANCE OF LEGITIMACY IN HIGHER EDUCATION ASSESSMENT

A PREPRINT

Jasper Roe [1]*, Mike Perkins [2], Peter Bannister [3], Leon Furze [4], James Wood [1],

[1] Durham University, United Kingdom

[2] British University Vietnam, Vietnam

[3] International University of La Rioja, Spain

[4] Deakin University, Australia

* Corresponding Author: jasper.j.roe@durham.ac.uk

May 2026

## Abstract

Artificial intelligence (AI) has disrupted assessment in higher education and accelerated the existing cycle of compounding performances. Institutional policies demand the demonstration of independent authorship, and commercial AI-enabled services enable students to feign independent thought and writing. This has led to enhanced institutional surveillance, such as AI detectors, which are subsequently circumvented using other technologies.

AI humanizers (internet-based services that alter AI-generated text to avoid automated or human detection) are a recent symptom in this performative cycle. Little is known about how these services operate and appeal to users, and their implications for educational assessment and integrity. Despite evidence suggesting that humanizers are rapidly increasing in popularity, they have attracted little scholarly attention to date. This paper presents the first exploratory, systematic investigation of AI humanizer websites, framed through Goffman's (1959) sociological account of dramaturgy. Using a systematic search and custom rubric, we cataloged 55 humanizer sites, assessing their performance of identity and conducting an in-depth multimodal critical discourse analysis (MCDA) of a purposive sample of three sites. Findings reveal that humanizers are readily available, offer free and premium paid services, and appear to perform similar functions. This includes the deletion and discursive absence of misconduct, the framing of AI humanization as a rational and defensible response to surveillance and flawed detection, and the appeal to mystification through advanced technology and implied endorsement by universities and corporations. We argue that humanizer services should be viewed as a diagnostic signal; they are a legible node in a feedback loop of performative assessment. We conclude that disrupting this cycle must be achieved through structural assessment reform, rather than technological solutionism.

# Introduction

This study approaches AI humanizers as symptomatic evidence of a deeper structural condition in higher education. Humanizers are best understood as the most recently visible node in a feedback loop in which assessment, student behavior, and commercial services are mutually implicated: institutional demands for a performance of independent authorship generate a market for AI tools to produce that performance; those tools generate institutional surveillance; and that surveillance generates further commercial services designed to conceal it. Documenting these services yields not only an account of the tools themselves but also a reading of the broader commercial architecture they inhabit and, in their very exposure, reveal.

The 2022 release of large language models (LLMs) and generative AI (GenAI) models has created an explosion of interest and controversy in educational contexts. The capabilities of GenAI have disrupted educational assessment by allowing users to create sophisticated textual and multimodal outputs with minimal effort or expertise. In other words, students can avoid demonstrating their learning in written assessments by offloading cognitive labor to these services. A fierce debate on how to reform, redefine, or adapt assessment in light of this disruption is ongoing; however, no clear consensus has emerged. At the same time, new services that further complexify assessment practices have appeared. These include GenAI text detectors and services for evading such detectors.

The detection landscape is itself shifting. A growing number of institutions have chosen not to adopt AI detection tools or have withdrawn from their use in response to documented concerns about reliability, bias, and procedural fairness (Liang et al., 2023; Weber-Wulff et al., 2023). If institutional detection recedes further, the market rationale for detection-evasion services may narrow. However, the analytical significance of humanizers does not rest solely on their role as evasion tools. As the most recently visible layer of a longer commercial integrity ecosystem, they expose the rhetorical strategies through which commercial actors, including, in a less concentrated form, the major providers of GenAI, frame the outsourcing of cognitive labor as rational, legitimate, and even unremarkable student behavior. This argument remains regardless of the trajectory of institutional detection policy.

Before the widespread adoption of GenAI, institutions had already come to rely heavily on similarity-matching systems, such as Turnitin, to identify copied or closely duplicated text. In response, students could turn to automated paraphrasing tools (APTs), or ‘text spinners’, which reworked source material into superficially new wording that was less vulnerable to text matching. Earlier studies of these tools have shown that they were designed to evade word-matching software rather than produce genuinely competent prose, often generating the kind of incoherent ‘word salad’ that itself became a marker of misuse (Prentice & Kinden, 2018; Roe & Perkins, 2022; Rogerson & McCarthy, 2017). In contrast, LLMs have enabled the production of fluent, original-seeming text at scale, while commercial AI detectors have been rapidly adopted as a new layer of surveillance intended to identify such use. However, these detectors have been widely criticized for weak reliability, bias, and vulnerability to relatively simple evasion techniques, raising concerns about fairness and procedural robustness in educational settings (Liang et al., 2023; Weber-Wulff et al., 2023; Perkins et al., 2024; Bassett et al., 2026; Giray, 2026).

Humanizers represent a simple evasion technique. These tools claim to transform AI-generated text into ‘human’ writing through stylistic edits that provide a more natural, ‘authentic’ voice. This is achieved through synonym substitution, sentence structure alteration, and simplification of semantic meaning using alternate wording (Giray, 2026). To examine the accessibility, features, and discursive strategies of humanizer sites as a symptom of this cycle, we conducted a critical, interpretivist investigation, pursuing the following research questions:

RQ1: What AI humanizer tools are currently available to students, and what dramaturgical features are deployed in their front-stage performance?

RQ2: Through what multimodal and discursive strategies do AI humanizer websites perform legitimacy work and conceal their backstage operations?

By answering these questions, we explore the availability, features, and strategies used by these services through the theoretical lens of dramaturgy. Specifically, we operationalize the concepts of a 'front stage' in which a carefully constructed performance is curated; 'back stage' in which the reality and underlying architecture of the performance are obscured; and 'leakage' in which the 'mask' of performance slips to reveal information which is intended to be concealed (Ekman & Friesen, 1969).

We build on Macfarlane's (2015) work, who posits that university education has shifted toward a Goffmanian public performance, in which some forms of assessment, such as reflective writing, serve as a stage for learners to conform to expected rules of production. Although not referencing Goffman specifically, Dawson (2020) has described the concept of 'security theater' to define approaches in assessment that appear to combat cheating but are ineffectual. In operationalizing our concept of performance, we take this notion of theater further by viewing unsecured written assessments specifically as often requiring a front-stage performance, in which student-authors are required to submit a demonstration of learning to an evaluating audience. In the era of GenAI, this performance may carry a specific expectation of producing an independent piece of work that is independently authored without GenAI text.

The dramaturgical lens is especially suited to humanizer platforms: unlike more transactional misconduct services, humanizers must actively perform legitimacy to a user who is already, at some level, aware of the ethical status of what they are accessing. Front stage, back stage, and leakage map directly onto this discursive work. Multimodal Critical Discourse Analysis (MCDA) then provides the methodological counterpart: while dramaturgy identifies the staging logic, MCDA enables a close analysis of how that logic operates simultaneously across language, image, and spatial composition, modes that a solely linguistic analysis would miss.

Furthermore, services which enable this performance are themselves engaged in a performance of their own. Humanizer sites must construct a convincing front stage that captures the attention of their audiences. In this way, GenAI detectors and AI humanizers become nodes in a feedback loop of performances. Just as students perform independence for an evaluating institutional audience, humanizers perform trustworthiness, rationality, and legitimacy to students.

We make three contributions to the higher education literature. First, we provide the first systematic empirical account of AI humanizer platforms, documenting the discursive and visual strategies through which these services normalize detection evasion and position themselves as rational tools for students navigating an asymmetric surveillance environment. Second, by applying Goffman's (1959) dramaturgical framework in combination with MCDA, we offer a methodological template for analyzing the performative dimensions of commercial misconduct services that extends beyond this specific phenomenon. Third, and most significantly, we argue that humanizer platforms point to a broader structural condition: the performative logic in which assessment demands, student behavior, and commercial services are mutually implicated.

# Literature Review

The scholarly response to GenAI and authorship in higher education has been substantial, yielding significant advances on the fallibility of AI detection tools (Bassett et al., 2026; Liang et al., 2023; Weber-Wulff et al., 2023), GenAI-conscious assessment design (Bearman et al., 2024; Corbin, Dawson et al., 2025; Fawns et al., 2025; Furze et al., 2024; Perkins et al., 2024), and a fundamental reframing of academic integrity around questions of assessment validity rather than misconduct (Corbin et al., 2026; Dawson et al., 2024). However, it has proceeded from a single analytical vantage point. The institution has been taken as the primary architect and communicator of the student informational environment, a position reinforced even in the field's most comprehensive collective statement of its own research priorities (Corbin, Bearman et al., 2026), which does not explicitly acknowledge the broader commercial digital environment students actually inhabit. This pattern extends to institutional policy documentation, which carries 'critical silences' (Luo, 2024, p. 651) that fail to account for the differing implications of GenAI use across diverse student cohorts (Bannister et al., 2024; Eaton, 2025; Perkins & Roe, 2024). Students, however, are not passive recipients of institutional messaging but active agents who reconstruct and reinterpret the information they encounter (Bearman et al., 2026; Wood & Pitt, 2025), and most spend more time on their screens (Ofcom, 2024) than they do on their studies (Advance HE & HEPI, 2025).

Beyond the institution, promotional information available to students on social media, through demographically targeted search engine advertising, and on company websites constitutes a parallel discursive environment in which students are simultaneously immersed. Media and gray literature narratives characterizing GenAI as an existential threat to higher education (Bearman et al., 2023; Jensen et al., 2025; Roe & Perkins, 2023) were consumed not only by educators but also by commercial actors structurally oriented toward extracting value from precisely such moments of institutional disruption (Komljenovic, 2021), and by students who arrived at their own conclusions about institutional authority, detection, and risk long before any formal guidance reached them.

AI humanizers join a lineage of commercial academic misconduct ventures, essay mills (Newton, 2018), and contract cheating platforms (Rowland et al., 2018), which exploit the fault lines between policy and practice (Bannister & Carver, 2025; Bretag et al., 2019) to entice students into parting with their money. AI has driven students to make complex boundary decisions without institutional support (Walton et al., 2025) in an asymmetric surveillance dynamic in which students are required to declare AI use, while institutions offer no equivalent transparency about how that declaration affects assessment outcomes (Luo & Dawson, 2025).

Similar studies have explored how academic misconduct ventures are marketed to students. Rowland et al.'s (2018) landmark analysis of contract-cheating websites documented the rhetorical strategies through which such services solicited student custom; however, that analysis necessarily confined itself to the linguistic surface of a multimodal object. For a student weighing detection risk against an imminent deadline (Bannister, 2026; Studiosity & YouGov, 2026), what ultimately dissolves hesitation is the coordinated operation of visual credibility (Saoula et al., 2023), social proof (Huang et al., 2025), and the spatial and compositional logic of the page itself (Leiser & Santos, 2024). The choreography of legitimacy is executed by borrowing the visual and discursive markers of trustworthy services (Ledin & Machin, 2019) to construct an appearance of institutional alignment that the platform's actual purpose could not sustain (Ho, 2021).

On the landing page, students who arrive carrying real pressures — deadlines, detection risk, and institutional distrust — find those pressures not merely exploited but reorganized into a story about themselves (Kjellgren et al., 2022), one that has already cast the institution as an adversary and the platform as a self-appointed savior (Medway et al., 2018; Tang et al., 2026). The recognitive experience (cf. Corbin et al., 2026; Wood & Pitt, 2025) of seeing their own situation written back at them lowers the guard (Shen, 2010) and folds their reservations into the narrative. The specter of solidarity masks an operation that ventriloquizes grievance against the institution (Crook & Nixon, 2021) while monetizing the desperation it performs sympathy with, extending the logic of predatory inclusion (Dawson, 2024). Beneath the transactional arrangement, the student consents to their own disempowerment dressed as liberation (Graef & Bostoen, 2026); the power governing their situation is unaccountable, unregulated, and invisible by design (Draper & Newton, 2017; Gaumann & Veale, 2024).

Students drawn into this cycle are not exercising unconstrained agency but are responding to a commercial architecture with no structural incentive to release them (Gray, 2022). These platforms assume the role of counter-epistemic authority, with non-institutional actors claiming a guidance function that displaces established institutional sources (Bartsch et al., 2025), while student engagement generates behavioral data of considerably greater commercial value than any subscription fee tendered (Zuboff, 2019). The epistemic, financial, and academic harms that follow are consequential, yet insufficiently examined within a field that has otherwise responded to GenAI with considerable urgency. This study seeks to broaden the field's analytical gaze to offer a more complete account of the full range of integrity-relevant messages students encounter, both institutional and commercial.

## Methodology

We view humanizer websites as socially constructed artifacts that produce meaning through a mixture of textual discourse and semiotic resources. This interplay is what we frame as dramaturgical. Goffman's (1959) work was originally focused on human interaction in social contexts, rather than online interactions or non-human actors. However, the dramaturgical dimension of the presentation of self has been studied on the Internet in the fields of social networking (Bullingham & Vasconcelos, 2013; Hogan, 2010; Whitty, 2008). This highlights the applicability of the framework for analyzing digital contexts.

In applying dramaturgical analysis to humanizer sites, we view the time from when the audience accesses the site to when they exit it as a bounded performance. The audience-accessible pages comprise the front stage, while the technology and owners of the sites are backstage. To develop these dramaturgical dimensions into an analytical framework, we created a scoring rubric that maps criteria onto a component of front-stage performance. We opted to use the constructs of visibility and reach (which relates to the way the performance reaches its audience), persona (the identity the site constructs for both itself and the audience), trust and authority artifacts (the props used to enhance the credibility of the performance), and mystification (the techniques used to enhance the performance and prevent backstage leakage). Each criterion was scored between 1 and 5, as shown in Table 1.

**Table 1: Scoring Rubric**

| Criterion | Theoretical Focus | Description | Scoring Guidance (1-5) |
|---|---|---|---|
| Visibility and Reach | The methods by which performance reaches the audience. | Measures how likely students are to encounter this tool through typical search behavior. This dimension ensures our analysis focuses on tools that shape student experience. Search engine positioning reflects market presence and investment in reaching student audiences. | 5 = Appears in the first ten results for Google. Clear paid advertising presence and/or social media visibility. |
| | | | 4 = Within the first page of Google results but not necessarily within the first ten results. |
| | | | 3 = Appears within the first or early second page of Google. |

| Criterion | Theoretical Focus | Description | Scoring Guidance (1-5) |
|---|---|---|---|
| | | | 2 = Appears in the second page of Google. |
| | | | 1 = Appearstowards the bottomof the second page. |
| Persona | Goffman (1959): Front stage performance and audience construction | This examines how the site constructs a student-user identity and develops its own identity. This is the establishment of who the performer is, and who the performance is for. | 5 = Has a clear orientationtowards students, referencing essays, professors, university, or similar lexical items. Includes student testimonialsor clear identification with common anxieties (e.g. deadlines). |
| | | | 4 = Strong but potentially implicit targeting of students through academic context. Users positioned in academic assessment situations. |
| | | | 3 = A mixed or ambiguous persona, with some references to students but not as the core target audience. |
| | | | 2 = Targeted mainlytowards ‘contentcreators’ or other audiences with little or indirect mention of students or educational contexts. |
| | | | 1 = Generic target of ‘writer’ with no mention of educational contexts or students. |
| Trust & Authority Artifacts | Goffman (1959): Props, setting, and impression management | Analyses the deploymentof visual and textual elements (Goffman's ‘props’) that construct legitimacy, safety, and authority. These artifacts comprise the front stage and indicate the depth of dramaturgical labor invested in the performance, | 5 = Trust artifacts are prevalent: shields, locks, detector logos, certificates or other appeals to authority. Professional and sophisticated design suggesting legitimacy. |
| | | | 4 = 3 – 4 types of legitimacy markers, for example detector screenshots or guarantees, professional design hinting at reliability. |
| | | | 3 = Moderate trust building with basic testimonials and some use of trust-building ‘props’ or artefacts. |
| | | | 2 = Minimal use of artifacts or props, maximum of one basic testimonial or iconography implying trust and quality, less polished web design. |
| | | | 1 = Plain, simplistic design with no visible trust artifacts or legitimacy markers; may appear poorly designed or non-functional. |
| Mystification; prevention of performance breakdown or backstage leakage | Goffman (1959): Mystification as dramaturgical strategy | Assesses the use of technical language and complexity to maintain the secrecy of the service’s function, prevent audience scrutiny, and obscure backstage" operations. | 5 = Mystification takes a central role, with the use of opaque, technology-heavy descriptions to confer legitimacy, for example, advanced algorithms (with little explanation). Function obscured through complexity. |
| | | | 4 = Significant technical jargon, with language implying the sophisticated use of unique technologies.. |
| | | | 3 = Moderate use of technical language with both plain-terms or accessible description of the technology with some jargon. Partially mystified description. |
| | | | 2 = Minimal jargon and technical terms, mostly accessible language. Minimal jargon: mostly accessible language with transparent function. |
| | | | 1 = Clear, plain language with little to no mystification. Direct and simple descriptions. |

Initial search plans included using multiple search engines; however, the high potential for reduplication of results led us to abandon this approach. To achieve ecological validity, we decided to focus solely on Google as our search engine for humanizer site cataloging. The rationale for this is that, according to recent statistics from SimilarWeb (2026), Google Search is the most visited website in the world and accounts for 90% of the search engine market share (StatCounter, 2026). This likely means that most searches for humanizer sites will occur through Google.

After the initial pilot searching, we developed a systematic search strategy. The keywords were subdivided by the researchers and are presented in Table 2, and the inclusion and exclusion criteria are presented in Table 3.

**Table 2: Keyword search terms**

| Search Term |
|---|
| AI humanizer/izer |
| Humanise/ize AI Text |
| Bypass AI Detection/Detector |
| Make AI Undetectable |
| Avoid AI Detection |
| Trick AI Detector |
| Pass Turnitin AI Detector |
| Pass Copyleaks AI detector |
| Pass GPTZero AI Detector |
| Pass Turnitin AI Detector |

**Table 3: Inclusion and exclusion criteria**

| Criterion Type | Inclusion Criteria | Exclusion Criteria |
|---|---|---|
| Functionality | The site offers a usable AI detection/humanizing service | Purely a blog/article/review |
| Access | Service is accessible, freemium, demo-based, or subscription | Requires enterprise onboarding with no public description |
| Relevance | Service claims relevance to authorship, originality, AI detection | Tool is only SEO filler or a generic writing assistant |
| Language | English-language OR English interface available | No English interface and no English description |
| Duplication | Unique service | Mirror site, affiliate branding, reseller site |

## Pilot Study

To enhance the application of the rubric, we engaged in an independent analysis of three websites selected by the lead researcher. We calibrated our analysis using inter-rater agreement, assessed using Krippendorff's alpha (Krippendorff, 2011). A result of 0.48 suggested moderate agreement among the research team. We did not designate the rubric as a positivist instrument for measurement but as a conceptual, structured framework to help guide analytical focus on shared dimensions. As a result, the focus on calculating Krippendorff's alpha was to identify any points of tension, divergence, or agreement in interpretation and improve our shared understanding of the criteria. Given the exploratory nature of the study and the interpretivist nature of our multimodal website analysis, this level of agreement was sufficient to move forward with our full analysis.

## Data Collection

Searches and collation of results occurred in December 2025 and January 2026. The research team conducted searches for allocated key terms in four geographical regions (Australia,

England, Spain, and Vietnam). Searches documented tool names, URLs, dates of access, search locations, and archiving. A copy of each site's homepage was saved for posterity. Each website was then independently scored using the rubric, and memos were recorded for further analysis. A total of 55 unique humanizer sites were collected. To enhance our exploration and analysis of front-stage website performance, we selected sites (N=3) for a detailed MCDA. These sites were selected as the highest-scoring on our rubric. The goal of this selection was not to be representative of all humanizer sites but to purposively sample the sites that appeared to engage most extensively with our criteria of performance.

**MCDA: Analytical Approach**

We employed MCDA following Machin and Mayr's (2012; 2023) approach to examine how AI humanizer websites construct and 'perform' legitimacy through text, image, and layout, in a bid to drive user engagement.

We take Fairclough's (2003) position that discourse analysis is not solely text but 'oscillates' between textual focus and the structuring of social practices. We analyze the lexical fields used to identify how word choices relate to particular topics, along with visual elements such as images, logos, and multimedia (Machin & Mayr, 2023), which we argue function as 'props' (Goffman, 1959) in the performance of site identity. We draw on Van Leeuwen's (2008) 'recontextualization' to explain how these sites operate discursively. Recontextualization involves the transformation of social practices as they move from one context to another; in this case, the transformation of inappropriately AI-generated text into the socially acceptable practice of 'writing improvement', 'humanization' or 'content enhancement.' Van Leeuwen (2008) identifies several key transformations that occur during recontextualization: substitution, deletion, rearrangement, and addition. We observe all four at work across humanizer websites.

Specifically, deletion operates through the systematic exclusion of terms like 'cheating,' 'plagiarism,' 'misconduct,' and so on, despite the fact that these words accurately describe the primary use case for many users. Substitution replaces these with euphemistic alternatives: 'humanize,' 'enhance,' 'refine,' and 'improve flow.' Addition introduces new legitimating discourses, particularly around writing quality, language barriers, and professional content creation, that were absent from the original practice. Rearrangement restructures the purpose hierarchy by foregrounding 'writing improvement' while backgrounding detection evasion, even when the latter is clearly the core function.

Kress and Van Leeuwen's (2006) work on visual design provides additional analytical framing, particularly through their framework for compositional meaning. Their three metafunctions (representational, interactive, and compositional) allow for the examination of how visual elements structure meaning-making across these sites. The concept of visual salience (what is made prominent through size, color, placement, and contrast) is particularly relevant to understanding how these sites foreground certain meanings while backgrounding others: a parallel to Van Leeuwen's (2008) recontextualization. Kress and Van Leeuwen's (2006) concepts of information value (the meaningful placement of elements in left/right, top/bottom, center/margin compositions), framing (how elements are connected or disconnected), and salience (how attention is directed) allow us to decode the visual grammar through which these sites construct their performances. Finally, we further integrate Goffman's (1959) dramaturgy by examining how 'front stage' legitimacy performances crack to reveal 'backstage' aims of integrity breaches. Using Goffman's (1959) dimensions, which underpinned our initial data charting, we developed an MCDA rubric to examine in detail how these dimensions of visibility and reach, persona, trust artifacts, mystification, and front and backstage were operationalized in our selected sites (those that scored the highest on the original rubric). We also sought evidence of leakage, which we define as errors or inconsistencies that disrupt the performance of the site. Leakage is a fundamental part of the study of deception and can be understood as a 'crack in the wall of secrecy' or an 'escape of direct indicators of the truth' (Gibson, 2014, p. 283).

**Table 4: MCDA Rubric**

| Rubric Dimension | Aspect of MCDA | Guiding question |
| --- | --- | --- |
| Visibility & Reach | Context analysis of where humanizer sites appear | How do these websites circulate, and how widely are they available? |
| Persona | Lexical field and discursive scripts | How is language used to create identities and foreground topics? |
| Trust Artifacts | Semiotic choice and icon analysis | What visual and textual 'props' are used to structure the performance of humanizer sites? |
| Mystification | Lexical field and discursive scripts | What use of jargon, technical terminology or 'trade secrets' are used to perform identity? |
| Front/Backstage | Inclusion and exclusion analysis | Are there any signs of 'leakage' between the front and backstage? |

# Findings

All sites demonstrated similar tensions between front- and back-stage operations. These are summarized in Tables 5, 6, and 7 below. The sites are referred to numerically, and their names or URLs are not included. The analytical focus of our study is on dramaturgical strategies across humanizer platforms rather than on an individual site. Furthermore, anonymization is used to reduce legal risk and ensure that sites are not promoted or advertised through our work.

**Table 5: Trust & Authority Artifacts across sites (N=3)**

| Artifact Type | Site 1 | Site 2 | Site 3 |
| --- | --- | --- | --- |
| Corporate logos | Present | Absent | Present |
| University logos | Present | Absent | Present |
| Detector logos | Present | Present | Present |
| Detector screenshots | Present | Present | Present |
| User statistics | Present | Present | Present |
| Satisfaction ratings | Present | Present | Present |
| Testimonials | Absent | Present | Present |
| Social media presence | Present (Inactive) | Present (Active) | Present (fragmented) |

**Table 6: Mystification strategies across sites (N=3)**

| Strategy | Site 1 | Site 2 | Site 3 |
| --- | --- | --- | --- |
| Algorithm references | Present - NLP and Deep Learning | Present - Advanced Algorithms | Present - Trillions of parameters |
| Scale claims | Present - 1 billion words/month | Absent | Present - Trillions of pieces of user data |
| One-click framing | Present | Present | Present |
| Success rate claims | 99.8% | 99%+ | 100% human |

| Strategy | Site 1 | Site 2 | Site 3 |
|---|---|---|---|
| Technical mode options | Present - Mode 1.0/2.0 | Present - Tuned for Turnitin, ZeroGPT, GPTZero | Present - Fast/creative/enhanced |

### Table 7: Front/Backstage leakage across sites (N=3)

| Indication of Leakage | Site 1 | Site 2 | Site 3 |
|---|---|---|---|
| Unendorsed logos | Present | Absent | Present |
| Inactive social media | Present | Absent | Present |
| Explicit misconduct admission | Absent | Partial (Face obscured in testimonials) | Present (student testimonial) |
| Concealed ownership/operation | Present | Present | Present |

### Website 1

The first website selected scored highly (5) on visibility and reach. On initial searches, it continued to appear on the first page of Google for multiple keywords. This suggests that the inclusion of this webpage has high ecological validity, as it is likely to be one of the first that potential users come across when looking for an AI humanizer. Although we cannot know whether the self-reported claims are truthful, the site claims at multiple points to have a large number of unique users. For a detection tool, the site reports more than 800,000 unique users, whereas for the humanizer, close to 3 million users are referenced.

The first platform that we selected describes itself not only in terms of academic assessments but also seems to appeal to a wide range of users in multiple contexts, referencing SEO specialists, marketers, businesspeople, and web designers among the common end users. To this end, the persona that the site seems to adopt is one that helps to optimize content from a strategic perspective, enhancing text to maintain a 'human' message while simultaneously avoiding detection. At the same time, there seems to be an impression that this site is a major producer of high-volume output, claiming to humanize over 1 billion words of text a month, and with consistent references to search engine optimization, search engine rankings, and a section in the FAQ specifically focusing on search engines. This helps to develop a persona of a well-established, professional service that appears to be well-used and adopted by large corporations. This business-like demeanor is furthered by the use of imagery: logos for the companies Adobe and Atlassian are featured on the front page of the site, just below the text entry boxes where users can enter text to check for AI traces and proceed to humanize their work. At the same time, these logos are not supported by text explaining why they are there; there is no reference to companies endorsing, using, or supporting the humanizing tool themselves. These 'props' appear to enhance the legitimacy of the site but do not have any 'backstage' components. The businesslike persona feeds into what appears to be a secondary audience: students and academics.

Interestingly, the website also repeatedly mentions bypassing AI detectors, including those that are most common in academic contexts, for example, Turnitin. Additionally, there are mentions of academic professionals and students needing to avoid AI detection systems. This leads to an overall less clear persona than in Website 2.

In terms of mystification, we noted the employment of many linguistic resources to obscure the actual process of humanization. These included 'state-of-the-art natural language processing algorithms' and 'deep learning models'. While it is understandable that proprietary tools would not be discussed in detail, the surface nature of these descriptions provides little to no insight into how the rewriting process works. Instead, the humanizing process is detailed as a simple input-output activity in which users post a text, and it is immediately humanized. Mystification is further solidified by the repeated mention of a 99% success rate in evading detection, which serves to enhance confidence yet gives no indication of how or why such a rate of success is possible.

From the perspective of dramaturgy, we note that the ‘front stage’ of the website emphasizes a simple, effective, and accurate service, using legitimacy-orienting symbols that suggest links to academic and corporate institutions. The performance is sophisticated, sleek, and appears to be well-designed to lead the user to sign up for the service, offering free trials, a range of pricing tiers, and even an affiliate marketing program. Other ‘props’ include social-media links, which lead to underdeveloped or inactive pages. This suggests that the backstage may be less sophisticated than it appears. Symbols of legitimacy and words that encourage trust and confidence are widespread; however, transparency remains elusive. In summary, the site appears to be somewhat mysterious in its persona, simultaneously offering to serve distinct audiences.

**Website 2**

The second website that we analyzed constructed a far more student- and scholar-facing persona. The homepage of the website directly addressed students who are required to submit work through AI detector sites and displays logos from popular detection websites, including Turnitin and Originality.ai. The site promises to be an undetectable, trusted AI detection bypasser used by more than 10,000 students. Interestingly, the user base claimed is far smaller than that of the first site but identifies that students are the target market.

The site’s overall persona is supportive and student-facing, with a subheading explaining how ‘humanizers support students’. First, it recontextualizes humanizers as protection against false positives, suggesting that AI detection is inequitable and that using a humanizing tool is a rational, ethically defensible act. Second, it presents humanizing AI-generated text as a way to enhance efficiency. The service suggests that students otherwise spend hours ineffectively humanizing AI-generated texts. The site also constructs the student as someone who needs to operate efficiently, treating assessment as a task to be completed quickly while managing academic demands and using technology strategically to minimize the time burden.

Similar to the first website, there is a clear emphasis on mystification regarding how AI-generated texts are humanized. Among the descriptive terms used, the humanizer is described as a ‘cutting-edge undetectable AI tool’, which is powered by ‘advanced algorithms’. Humanization is again described as something that is enacted through the click of a single button; however, the underlying mechanism through which the transformation process takes place is concealed.

The website includes AI detector logos and seeks to build confidence and establish legitimacy by showing screenshots of AI detector result pages for seven different online detectors.

This site is also multimodally rich, embedding short YouTube clips which show students discussing the benefits of the tool and explaining their experience. The first video describes a narrative of a student who got caught and failed a course as a result of their use of AI in an essay; the second describes a student using a humanizer to submit an AI-generated essay close to a deadline; the third explains an experience of watching other, less-informed students getting caught for submitting poorly edited ChatGPT-generated essays. These ‘props’ are highly effective, mimicking common social media short-form video reels, which include an informal, narrative-like sales pitch drawing on ‘real’ student experience. This sophistication suggests a significant investment and development of the tool. At the same time, there are some hints that this tool is not entirely legitimate; while the first and last clips show the speaker’s faces, the second clip obscures the user's face, suggesting that they do not want to be identified for their use of a humanizer. The second page demonstrates a far more comprehensive social media presence, including a TikTok account with thousands of followers and close to 100,000 unique viewers. Taken together, these complementary elements contribute to a sense of student community, validation, and practical effectiveness. Overall, the site suggests a pragmatic and reasonable response to a set of challenging educational circumstances (time pressure, risk of false positives, and severe academic consequences).

**Website 3**

Similar to the first two humanizer sites, the third site also demonstrated a strong first-page presence across multiple searches. In this case, the site claimed to have 1.5 million users and to support over 400 companies, along with multiple logos from large international corporations. In addition, the site's landing page shared many similarities with the first two, claiming a significant number of users. Upon inspecting the social media presence of the third website, we noted evidence of significant social media impact on some sites, with Facebook being the strongest, with over 3,000 followers, and content in a range of languages suggesting an international reach. In contrast, although YouTube and TikTok profiles were embedded in the homepage of the website, they led to sparsely populated or empty pages, with little to no followers. This fragmented social media seems to be a hallmark of the sites examined.

Similar to Website 2, which used the logos of organizations and universities, Website 3 operates with a similar, yet expanded portfolio of logos. These include Johns Hopkins University, Princeton University, and many large international corporations, such as Netflix, Airbnb, and eBay. Graphically, the homepage appears polished, sleek, and sophisticated, with dynamic text and a moving set of adjectives: engaging, undetectable, and human. Similar to Websites 1 and 2, a text box that allows 'AI text' to be pasted in and humanized is present, with the option to select from different modes: fast, creative, and enhanced.

The third site presents a broad potential user base, although there are some suggestions that academic use is the primary purpose. The use of testimonials on the first page includes a moving carousel of small headshots of individuals, alongside their name, profession, and a short paragraph on using the service. Among the professions mentioned are writers, journalists, entrepreneurs, CEOs, and academics. Academic framing is foregrounded with a student testimonial that explicitly states that users' professors can identify AI-generated text and that the service allows the user to submit essays with more confidencethat they will not beg accused of using AI. In a similar vein to the first two sites, an FAQ section is available at the bottom of the homepage, which offers multiple leading questions (for example, what free AI humanizing service can I use?) and affirmative, enthusiastic responses (yes!) Overall, the tone is supportive, aiming to reduce the burden on the user and help 'do the heavy lifting' for the user. This is perhaps the area in which recontextualization is most clear; the site is positioned as a supporter and collaborator rather than a tool that enables potential misconduct.

The third site had the most sophisticated deployment of 'props' to perform legitimacy. This included logos from high-prestige universities (such as Princeton and Johns Hopkins), and corporate logos (Netflix, Airbnb). Much like the first site, these logos were unexplored, unexplained, and did not have a strong correlation or description of endorsement. Other trust and authority artifacts included the use of satisfaction ratings (4.9/5), and a carousel of screenshots of AI detectors being 'passed' using the service, along with green ticks. Pathway maps showing the process, including color shift from red to green mirror similar visuals in both sites 1 and 2, while the diverse carousel of different user types helps to build credibility.

Mystification was most clearly visible in site 3. The use of numbers to suggest complexity is employed, claiming '1.6 trillion parameters' were used in training the humanizer service. This evokes the scale of flagship large language models (such as OpenAI's GPT series); the meaning of the parameters is unclear, yet it is a persuasive prop in developing the front stage of a professional, technically sophisticated service that uses high-end technology. This is furthered with claims similar to those found on the first two sites, such as 'state-of-the-art' technology and training on 'trillions' of pieces of content. Multiple modes, such as fast, creative, and enhanced contribute to mystifying the process without explaining what exactly the different modes do to the service. Finally, the suggestion of multilingual uses provides support for the idea of an international business, yet the data are inconsistent: the service claims both 30 languages and 50 languages are available.

Returning to the concept of the site as a front stage, in which a carefully developed performance takes place, the third site appears not to maintain a mask of quasi-legitimacy but explicitly acknowledges the goal of humanizing; testimonials explain how students can confidently submit AI-generated work as their own. At the same time, the use of social-media logos suggests an active online social community; however, in reality, these are essentially inactive.

## Discussion

These sites are underpinned by a similar logic: they recontextualize AI-generated text and evasion of detection through humanization as rational, pragmatic, and acceptable behaviors. These three sites appear sophisticated, well-designed, and polished, drawing on similar practices of legitimacy, such as testimonials, claims about the number of users, declarations of undetectability, and sophisticated, cutting-edge technology. At the same time, these sites engage in mystification practices, with none explaining exactly which underlying architectures or techniques are applied to humanize text.

Goffman (1959, p. 36) argued that the 'impression of reality' which is created through performance is fragile, and errors must be corrected before the performance takes place (p. 27). We noted that certain elements of the polished, professional, established, and supportive service that these humanizers offer are fragile, and errors that hide just beneath the surface point to the limits of the performance. This 'leakage' results in the metaphorical mask slipping and impacts the performance. For example, despite the sites having pages for terms of service, contact forms, and in some cases, registered trading names, no site provided a registered business address. At the same time, the use of logos belonging to prestigious institutions and companies evokes trust, reliability, and quality at first glance; however, the lack of explanation or links to a source or explanation makes the connection appear questionable or tenuous. A final case of leakage occurs in social media icons, which in some cases lead to active accounts, but more often lead to blank or nonexistent pages. The use of social media iconography suggests an active community of user engagement, yet the reality is that this is a curated, performative act. We do not yet know how convincing these performances are to students as users.

The analysis of humanizers, as a marginal commercial phenomenon, says more about the ecosystem they occupy than about the platforms alone. Humanizer vendors cannot borrow institutional legitimacy in the same way major AI providers can; they must construct it entirely from front-stage materials, which is why the strategies are so visible here. The deletion of misconduct, rational self-defense framing, borrowed institutional iconography, and mystification through proprietary capability claims are not unique to this product category. In a softened form, these run through the marketing of mainstream AI providers as well. Humanizers are the concentrated version of these moves, the part of the iceberg above the surface; what they reveal below is worth the field's attention.

## Limitations

Although our study sheds light on the topic of humanizers and their impact on assessment in higher education, there are several limitations that must be acknowledged. First, the 'backstage' operation of humanizers is concealed, meaning that we have only been able to analyze what is publicly visible. We cannot be sure who is operating these sites, from where, and how they work. Second, our study did not seek to assess the efficacy of humanizer tools in bypassing AI detection technologies. While this would be a useful addition to the literature, it was not within the scope of our work but remains a key question for future empirical work. Third, while we sought to search for humanizers using Google, as it represented the best ecological validity available, we do not have data on how students are accessing these sites; it is possible that social media, word of mouth, or online messaging services contribute to the spread of these tools. In addition, humanizer sites appear to be constantly changing, updating, or disappearing. This makes follow-up studies difficult, as replicability is challenging. During our data collection, we noted several sites that disappeared and became inaccessible. Finally, although we have sought to utilize both a rigorous scoping-style analysis for our first research question, combined with a rich, interpretive analysis for our second, we note that our findings are not easily generalizable.

## Conclusion

This study explored the appearance of AI humanizers and their implications for educational assessment, framed through Goffman's (1959) dramaturgy in combination with MCDA. We identified that AI humanizer tools are readily available via the Internet and appear to operate within shared rules of engagement. The sites that we investigated constructed a seemingly sophisticated front-stage performance, deploying trust artifacts (such as university and corporate logos), mystification strategies (such as proprietary algorithms), and carefully curated personas that rationalize and justify their existence. Van Leeuwen's (2008) concept of recontextualization provided insight into the discursive construction of humanizers; the deletion of misconduct and the foregrounding of legitimacy and rationalization serve to persuade the user that such tools are irrelevant to academic integrity or the facilitation of misconduct.

Finally, we note that the emergence of humanizers is a structural feature of assessment in higher education. The significance of this study extends beyond humanizers themselves: as commercial services that must construct their legitimacy entirely from front-stage materials, they are a concentrated and legible version of rhetorical strategies that are increasingly ambient across the commercial AI-in-education landscape. This is a feedback loop in which assessments that demand a performative output generate a market for commercial AI tools to satisfy that demand; those tools generate institutional surveillance; and that surveillance generates a further commercial market for evasion. Commercial actors are not accelerants of this cycle — they are its architecture. From the major LLM providers who enable the initial AI-assisted writing to the detector vendors who sell institutional security and the humanizer platforms that market evasion, each layer of the cycle is inhabited by commercial interests with no structural incentive to interrupt it. The consequences of this situation are serious, causing multiple forms of harm, including epistemic, academic, and financial harm to students. The implications of our research suggest that the introduction of new tools may not disrupt the loop of performance in assessment, and the structural underpinnings of educational assessment require rethinking.

## Acknowledgements

The authors used GenAI tools to support specific aspects of the research and manuscript preparation, including assistance with editing and drafting some sections of text, which were subsequently revised; summarizing and paraphrasing content; providing feedback on drafts; and checking grammar. The tools used were Claude Pro (Opus and Sonnet, versions 4.2 and 4.6) and ChatGPT (GPT-5.5), selected for their capacity to provide sophisticated feedback on textual outputs. These tools were used in a supporting capacity and not to replace core author responsibilities. All AI-generated outputs were reviewed, verified, and refined by the authors, who take full responsibility for the final content.